\documentclass[reprint,secnumarabic,amssymb, nobibnotes, aps, prd,longbibliography]{revtex4-1}

\usepackage{graphicx} 
\usepackage{subfig,epstopdf}
\usepackage{natbib,url}
\usepackage{amsmath} 
\usepackage{color}

\newcommand*{\be}{\begin{equation}}
\newcommand*{\ee}{\end{equation}}

\def\begineq{\begin{equation}}
\def\endeq{\end{equation}}

\def\begineqn{\begin{equation*}}
\def\endeqn{\end{equation*}}

\def\beginar{\begin{eqnarray}}
\def\endar{\end{eqnarray}}
\def\beginarn{\begin{eqnarray*}}
\def\endarn{\end{eqnarray*}}

\def\Rat{\widetilde{Ra}}

\begin{document}

\title{The sensitivity of rapidly rotating Rayleigh--B\'enard convection to Ekman pumping }%

\author{Meredith Plumley$^1$}  \email[corresponding author: ]{meredith.plumley@colorado.edu} 
\author{Keith Julien$^1$}
\author{Philippe Marti$^{1,2}$}
\author{Stephan Stellmach$^3$}

\affiliation{
$^1$Department of Applied Mathematics, University of Colorado, Boulder, CO  80309, USA \\
$^2$Institute of Geophysics, ETH Zurich, Zurich 8092, Switzerland \\
$^3$Institut f\"ur Geophysik, Westf\"alische Wilhelms-Universit\"at, D-48149, M\"unster, Germany }

\begin{abstract}
The dependence of the heat transfer, as measured by the nondimensional Nusselt number $Nu$,  on Ekman pumping  for rapidly rotating Rayleigh--B\'enard convection in an infinite plane layer is examined for fluids with Prandtl number $Pr = 1$. A joint effort utilizing simulations from the Composite Non-hydrostatic Quasi-Geostrophic model (CNH-QGM) and direct numerical simulations (DNS) of the incompressible fluid equations has mapped a wide range of the Rayleigh number $Ra$ -- Ekman number $E$ parameter space within the geostrophic regime of rotating convection. Corroboration of the $Nu$--$Ra$ relation at $E = 10^{-7}$ from both methods along with higher $E$ covered by DNS and lower $E$ by the asymptotic model allows for this extensive range of the heat transfer results. For stress-free boundaries, the relation $Nu-1 \propto (Ra E^{4/3} )^{\alpha} $ has the dissipation-free scaling of $\alpha = 3/2$ for all $E \leq 10^{-7}$. This is directly related to a geostrophic turbulent  interior that throttles the heat transport supplied to the thermal boundary layers. For no-slip boundaries, the existence of ageostrophic viscous boundary layers and their associated Ekman pumping yields a more complex 2D surface in $Nu(E,Ra)$ parameter space.  For $E<10^{-7}$ results suggest that the surface can be expressed as $Nu-1 \propto (1+ P(E)) (Ra E^{4/3} )^{3/2}$ indicating the dissipation-free scaling law is enhanced by Ekman pumping by the multiplicative prefactor $(1+ P(E))$ where $P(E) \approx 5.97 E^{1/8}$. 
It follows for $E<10^{-7}$ that the geostrophic turbulent interior remains the flux bottleneck in rapidly rotating Rayleigh--B\'enard convection. For $E\sim10^{-7}$, where DNS and asymptotic simulations agree quantitatively,  it is found that the effects of Ekman pumping are sufficiently strong to influence the heat transport with diminished exponent $\alpha \approx 1.2$ and $Nu-1 \propto  (Ra E^{4/3} )^{1.2}$.

\end{abstract}

\maketitle

\section{Introduction}

Rotation and convection are key components of many geophysical and astrophysical systems, including planetary oceans, atmospheres and interiors, as well as stellar interiors \cite{jones2007thermal,jM99,mM05}.  These systems are typically rapidly rotating, highly turbulent and observationally remote, rendering them difficult to study.
To investigate the fundamental dynamics of these systems in a simplified setting many investigations employ the Rayleigh--B\'enard configuration of a fluid confined between two parallel rotating plates with an imposed buoyancy inducing temperature difference $\Delta T$. Complications are further reduced by assuming the rotation, at rate $\Omega$, is aligned with the gravitational $z$ axis and using a horizontally periodic Cartesian box for simulations \cite{Stellmach, Plumley} or an upright cylinder for experiments \cite{Cheng}, both with vertical scale $H$.  

The rotation of the system can be quantified by the Ekman number $E = \nu/2\Omega H^2$, which measures the importance of the Coriolis force relative to viscous dissipation.  The Rayleigh number $Ra = \alpha g \Delta T H^3/\nu \kappa$  measures the magnitude of thermal forcing relative to dissipative effects. Here $\nu$ is the kinematic viscosity, $\kappa$ is the thermal diffusivity, $\alpha$ is the thermal expansion coefficient and $g$ is gravitational acceleration. The rotational constraint of the system is determined by the Rossby number $Ro = U/2\Omega H$, a ratio of the strength of the Coriolis force to inertial forces. Here $U$ is a characteristic velocity scale, which when expressed as a free-fall velocity gives an \textit{a priori} measure $Ro = E \sqrt{Ra/Pr}$. The Prandtl number $Pr = \nu/\kappa$ is an attribute of the fluid; for air, $Pr \approx 1$.  These nondimensional parameters can be tied to the efficiency of the vertical heat transfer in the system through the Nusselt number $Nu = q H/\rho_0 c_p \kappa \Delta T$, where $q$ is the heat flux and $\rho_0 c_p$ is the volumetric heat capacity.

The interplay between rotation and buoyancy and the sensitivity of the flow to each of these forces can be examined through the heat transfer scalings of $Nu$ with each parameter. The precise nature of these scalings has become the focal point of both laboratory experiments and numerical simulations due to the relative ease of measuring the Nusselt number and the link between these scalings and the realized flow \cite{jmA15}. In fact, the power law scalings provide a proxy for determining the flow morphologies \cite{Julien12, Nieves} and once understood, can potentially be used to extrapolate to settings like planetary interiors.  Results from numerical simulations
have revealed the existence of four flow morphologies with increasing $Ra$ at fixed $E$. These include cellular motions that give way to convective Taylor columns formed out of the instability and synchronization of the thermal boundary layers. This gives way to
state of boundary layer plumes which synchronize intermittently. Eventually, the inability of the boundary layers to synchronize gives rise to a state of geostrophic turbulence (GT) in the fluid interior \cite{Sprague, Julien12, Nieves, Stellmach}.

 Assumptions for the functional form of the heat transport scaling law are not unique; one such consideration is
\be
Nu -1 \propto Pr^{\gamma} Ra^{\alpha} E^{\beta} \, ,
\label{E:oldNu}
\ee
allowing $\gamma$, $\alpha$ and  $\beta$ to independently determine the sensitivity to fluid type, thermal forcing and rotation, respectively. This choice follows logically from a common strategy within the rotating Rayleigh--B\'enard fluids community to either vary the rotation rate while holding $Ra$ fixed to focus on $\beta$ \cite[e.g.,][]{Kunnen16, Ecke14, SWeiss} 
or instead vary the heating at constant rotation to find $\alpha$ \cite[e.g.,][]{Stellmach}. This difference in approach has led to difficulties in consolidating results into a robust scaling theory as the connection between exponents $\alpha$ and $\beta$ remains ill-understood. 
Studies frequently report differing values of the exponents obtained from various slices through the multi-dimensional parameter space \cite{Cheng, Kunnen16}.  Accordingly, there has been much debate about the value of the scaling exponents within the field.
We proffer that this contention is formed from a lack of insight into the $Nu$ surface in $Ra$--$E$ parameter space and how the directional path of a given study within that parameter space affects the scaling exponents. Admittedly, the paucity of both laboratory and DNS data within the low $Ro$ regime constitutes the primary cause for this situation.

This issue has been resolved for the case of stress-free boundaries with many verifications of the dissipation-free scaling law of $\alpha = 3/2$, $\beta = 2$ (implying $\gamma = -1/2$) for a turbulent interior that throttles the efficiency of the heat transport through the boundary layer regions \cite{Julien12PRL, Stellmach, Kunnen16}. Thus 
\be
Nu -1  \propto Pr^{-1/2} Ra^{3/2} E^{2}
\label{E:SFnu} 
\ee
for stress-free boundaries. This relation can be further rearranged as
\be
Nu -1 \propto Pr^{-1/2} \Rat^{3/2}
\label{E:SFnu2}
\ee
illustrating that the heat transport is controlled solely by a supercriticality parameter through the reduced Rayleigh number $\Rat = Ra E^{4/3}$; a parameter that appears naturally in the asymptotic theory of rotating convection \cite{Chandrar}. More generally, a sole dependence of the scaling law (\ref{E:SFnu}) on $Pr$ and $\Rat$ implies $\beta = 4 \alpha /3 $, indicating the expectation that the heat transport law can be uncovered by measurement at either fixed $Ra$ or $E$.

We note specifically that  the $E^{4/3}$ Ekman dependence in the supercriticality parameter $\Rat$ arises as a consequence of critical Rayleigh number $Ra_{c}\sim E^{-4/3}$ for rotating convection \cite{Chandrar}, thus $\Rat \sim Ra/Ra_c \sim Ra E^{4/3}$. This should be 
juxtaposed with the non-rotating problem where $Ra_c^{nr}$ is a fixed constant which results in  a direct parameter-independent link between $Ra$ and the supercriticality parameter, such that $\Rat^{nr} \sim Ra/Ra_c^{nr} \sim Ra$.

For no-slip boundary conditions, \citet{King} first proposed a scaling law akin to the Malkus--Howard \cite{Malkus} theory requiring the thermal boundary layers to be marginally stable. Consistent with $\beta = 4 \alpha/3$, this produced scaling exponents $\alpha = 3$ and $\beta = 4$. Neither this result nor the stress-free result is observed \cite{Julien16, Plumley, Kunnen16, Cheng, jmA15} and it is now generally accepted that the presence of Ekman pumping associated with no-slip boundaries further complicates the heat transport scaling law. Progress towards an understanding of this fundamental interaction for the case of no-slip boundaries has been furthered through a collaborative effort combining heat transfer data from asymptotic theory \cite{Julien16}, direct numerical simulations \cite{Stellmach, Kunnen16}, and laboratory experiments \cite{Cheng, Ecke14}.

Here we present an empirical investigation of  the asymptotic heat transfer scaling appropriate for $Pr =1$ fluids with no-slip boundaries.  We focus on $Pr = 1$ fluids because it is known to rapidly enter the regime of geostrophic turbulence as a function of buoyancy forcing \cite{Julien12, Nieves}. Presently, it is the only case for which data from DNS and asymptotic models exist for the geostrophic turbulence regime. Laboratory experiments still trail in their ability to access this regime \cite{Ecke14, jmA15}. The asymptotic model,  referred to as the composite non-hydrostatic quasi-geostrophic model (CNH-QGM) is valid in the limit $E\downarrow 0$ \cite{Julien16}. It imposes 
a pointwise geostrophic balance where the Coriolis force is balanced by the pressure gradient force. Buoyantly driven fluid motions then evolve under the dominant action of horizontal inertial advection and horizontal dissipation. Currently, this model represents the only means of probing the turbulent -- low $Ro$ regime of rotating Rayleigh--B\'enard convection.

One aim of this work is to explore how different scalings can coexist within the complex $Nu$ parameter space for no-slip boundaries. This is addressed through an analysis of runs for both fixed $E$ and $Ra$ and the consolidation of these results to form a surface of $Nu(\widetilde{Ra},E)$ over a range of $ 8 < \widetilde{Ra} \leq 100$ and $ 10^{-11} \leq E \leq 10^{-5}$.  We utilize a combination of results obtained from DNS \cite{Stellmach, Kunnen16} at and above the current lower bound of $E \approx 10^{-7}$ and the asymptotic CNH-QGM for $E \leq 10^{-7}$ \cite{Julien16}. The fidelity of this model was verified with DNS at $E = 10^{-7}$  for $Pr = 1$ in \citet{Plumley}.  While visually helpful, the complexities of the surface render it difficult to characterize analytically. However, low Ekman results can be utilized to quantify the effect of pumping within the  GT regime. A key question we address is to what extent Ekman-pumping disrupts the ability of the geostrophic turbulent interior to throttle the heat flux through the layer. This may be understood by assessing the departure from the dissipation-free scaling law (\ref{E:SFnu2}). Inherent in this statement
is the need to produce a heat transport scaling law that provides an understanding of measurements obtained from different pathways through
$Ra$--$E$ parameter space. 



Based on the results of the asymptotic model at low $E$, we extend (\ref{E:SFnu2})  and propose the scaling law 
\be
Nu - 1  = c_1 (1 + P(E)) Pr^{\gamma} \Rat^{\alpha} 
\label{E:Nu}
\ee where $c_1 \approx 1/25$ is a constant factor \cite{Julien12PRL}. This states that the effect of Ekman pumping is to enhance the heat transfer by the multiplicative factor $(1+P(E))$. 

\section{Stress-free results}

\begin{figure}[h]
\includegraphics[width=.45\textwidth]{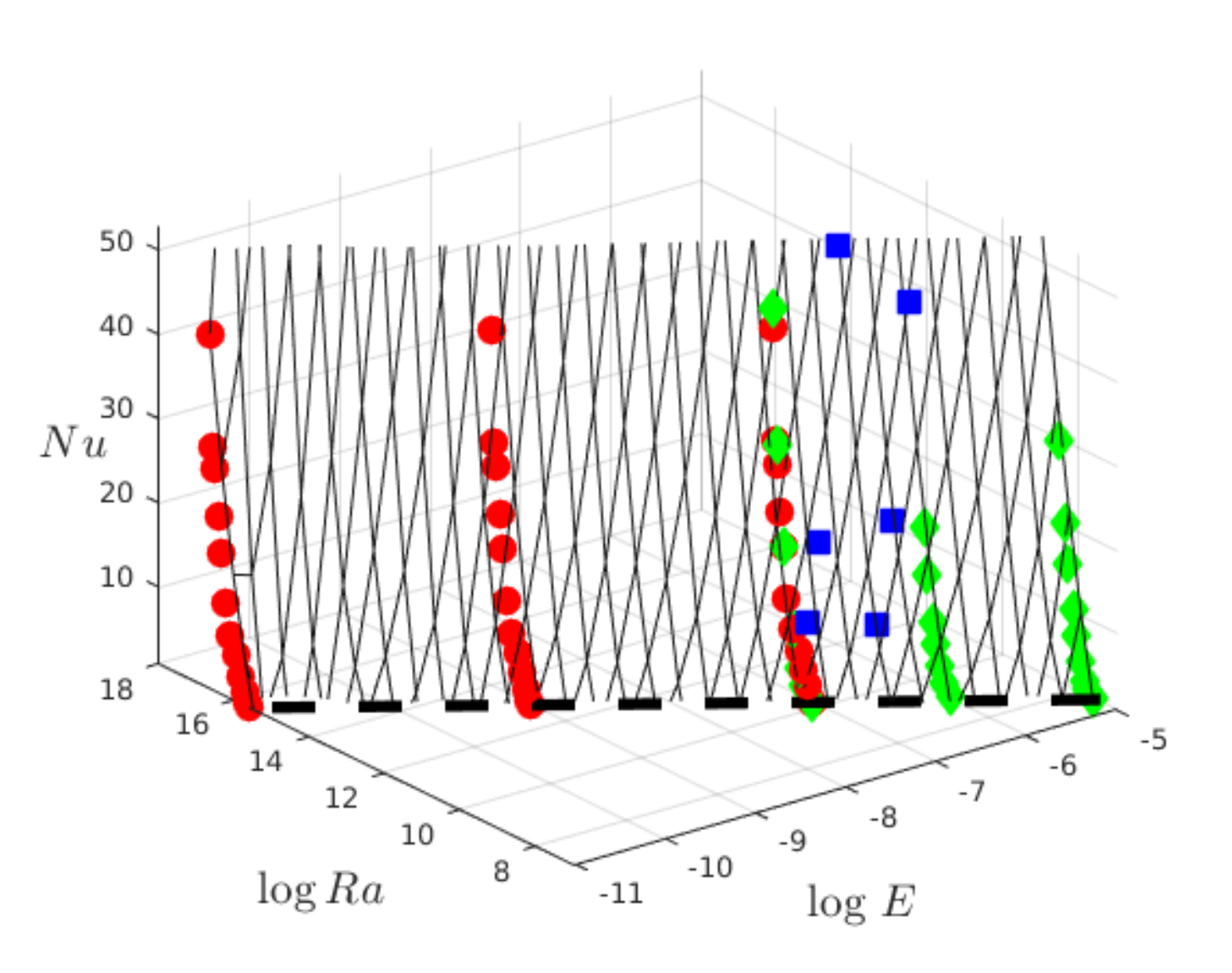} 
\caption{\small A 2D surface  of  $Nu$ as a function of $Ra$ and $E$ for stress-free boundaries and $Pr =1 $. Results marked by (red) circles denote data obtained by the CNH-QGM \cite{Sprague}.  DNS data for $E \ge 10^{-7}$ with $Ra$ constant (blue squares) and $E$ constant (green diamonds) are also included. The dashed line indicates $Ra_c  = 8.7 E^{-4/3}$.}
\label{F:surf_SF_RA}
\end{figure}

\begin{figure}[h]
\includegraphics[width=.45\textwidth]{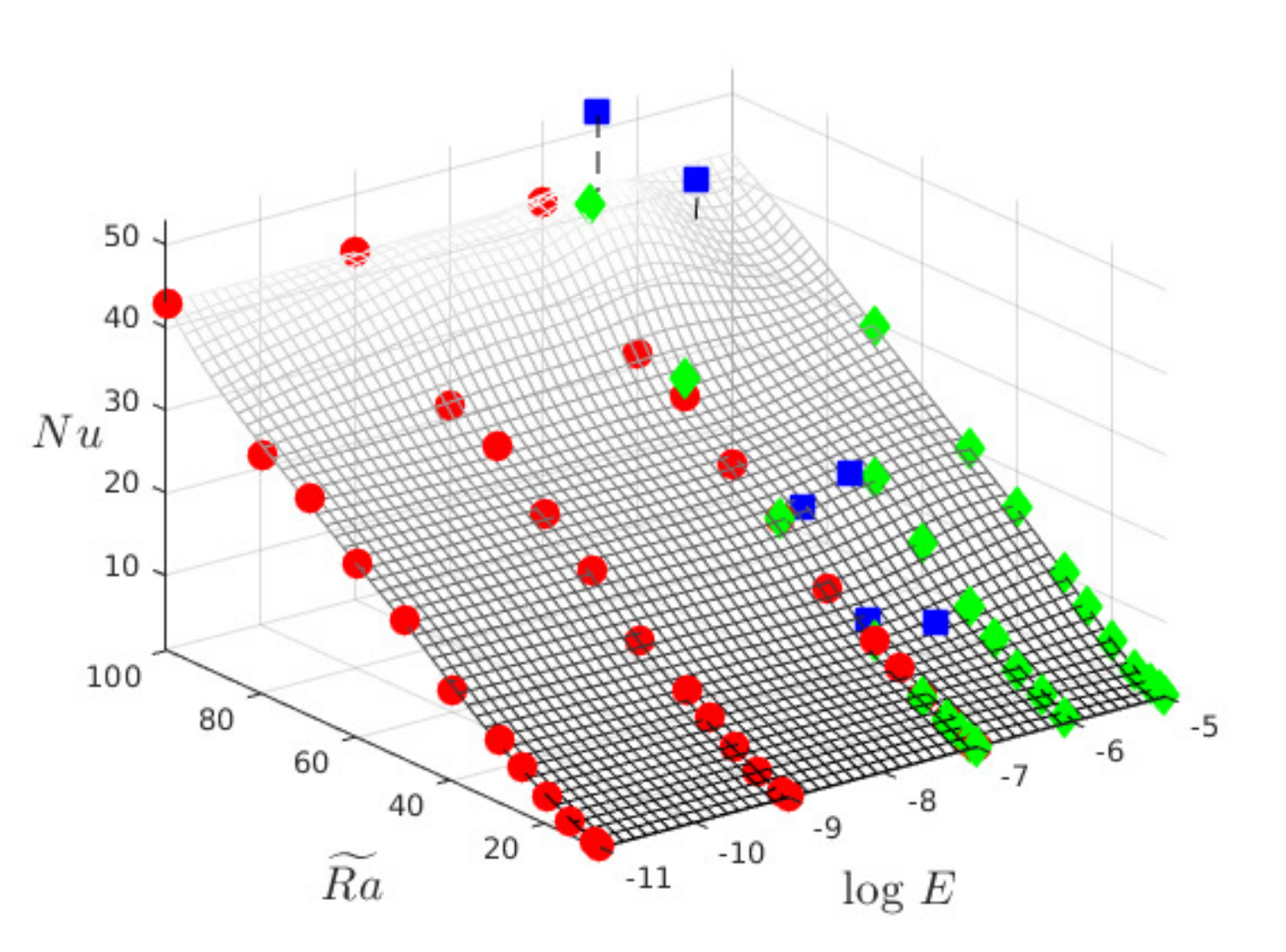} 
\caption{\small The 2D surface plot  of  $Nu$ as a function of $\Rat$ and $E$ for stress-free boundaries and $Pr =1 $. Results marked by (red) circles denote data obtained by the CNH-QGM \cite{Julien16}.  DNS data for $E \ge10^{-7}$ obtained from holding $Ra$ constant (blue squares) \cite{Kunnen16} and $E$ constant (green diamonds) \cite{Stellmach} are also included. Contours of the gridded surface for $E\le 10^{-7}$ are parallel to the $E$ axis indicating independence of $E$. Dashed lines from the squares indicate those points were not included to form the surface.}
\label{F:surf_SF}
\end{figure}

\begin{figure}[h]
\includegraphics[width=.45\textwidth]{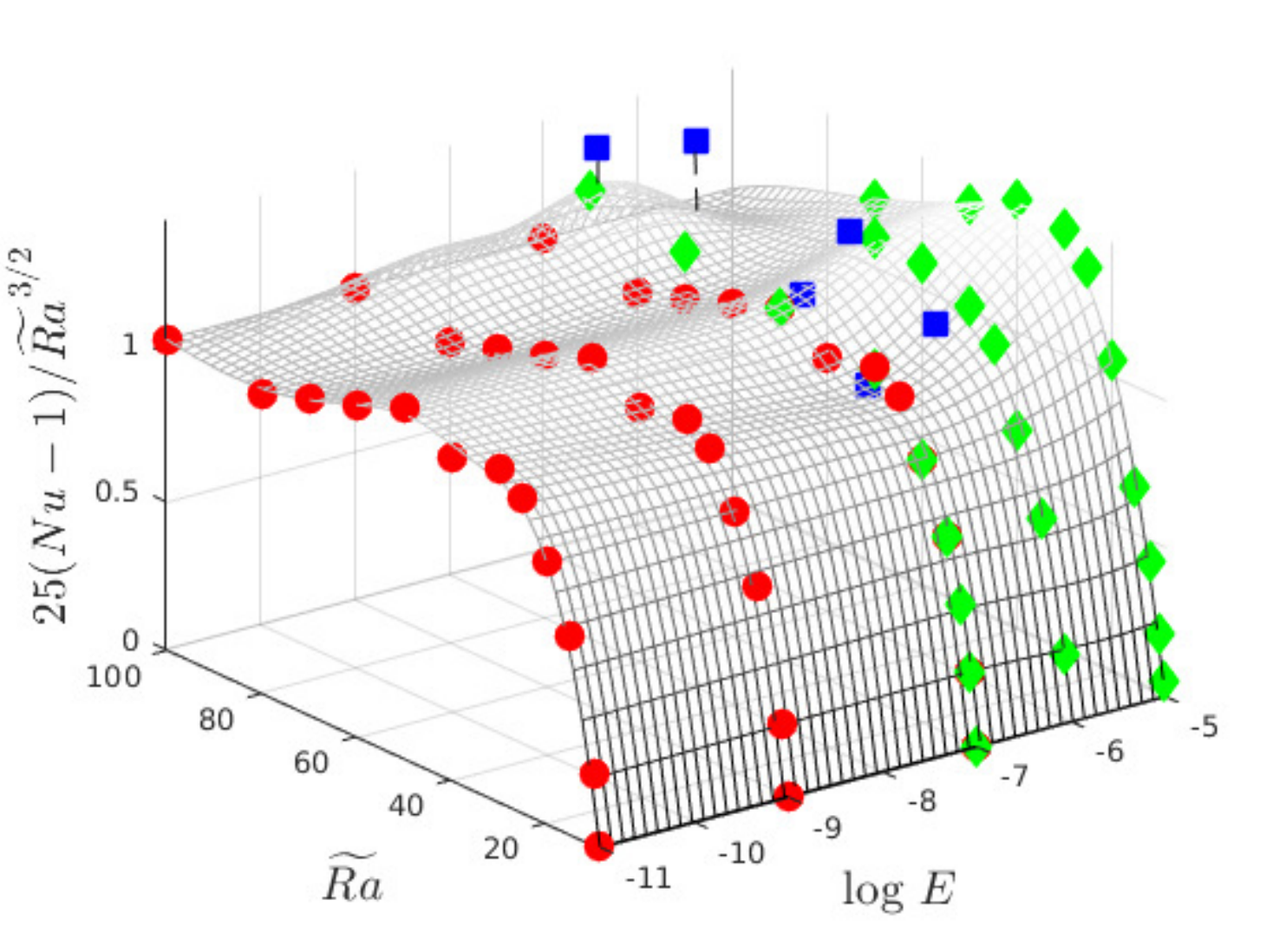} 
\caption{\small  Compensated surface plot  $25 (Nu-1)/\Rat^{3/2}$ for stress-free boundaries with $Pr = 1$. Results from the CNH-QGM for $E \leq 10^{-7}$ are denoted by circles (red), DNS at constant $E$ by diamonds (green), and DNS at constant $Ra$ by squares (blue). Dashed lines from the squares indicate those points were not included to form the surface.
}
\label{F:surf_SF_comp}
\end{figure}

In figure~\ref{F:surf_SF_RA} we illustrate the 2D surface obtained for $Nu$ in $Ra$--$E$ parameter space. 
The surfaces were generated from a combination of DNS and  CNH-QGM data and the MATLAB scattered\_interpolant and gridded\_interpolant functions. 
A coarse mesh surface is created from all the scattered data, where the scattered\_interpolant function generates a continuous surface of the given data points using linear interpolation. This surface is then smoothed using the spline option of the gridded\_interpolant function to reduce any artificial variations.

Evident from this surface is the delayed onset of convection for increasing rotation rates (decreasing $E$) where ${Ra_c} \sim E^{-4/3}$
for $Nu=1$ \cite{Chandrar} (see dashed line). Also observable is the excellent quantitative agreement between DNS and CNH-QGM data at $E=10^{-7}$.
However, we find
the $Nu$--$Ra$--$E$ surface of less visual utility in comparison to $Nu$--$\Rat$--$E$, which has the added benefit of providing equivalence in the supercriticality for differing $E$ (figure~\ref{F:surf_SF}).

As noted, in the case of stress-free boundaries, exponents  in  (\ref{E:SFnu2})  can be predicted  by seeking the turbulent dissipation-free scalings $\alpha=3/2$ and $\gamma=-1/2$, and from (\ref{E:oldNu}) it follows $\beta = 2$.  This scaling holds for all $E\leq 10^{-7}$ \cite{Julien12PRL}, thus creating a simple 2D surface in $Nu$--$\Rat$--$E$ parameter space (figure \ref{F:surf_SF}). 
For $E\le 10^{-7}$, it is observed that the contours are parallel to the $E$ axis confirming the appropriateness of the supercriticality parameter $\Rat$.
Departure from the planar surface is visible for $E > 10^{-7}$ which may be taken as evidence of  a breakdown in the asymptotic regime. 
The features of the $Nu$ surface are more pronounced in the compensated surface normalized by the scaling law
$c_1 \Rat^{3/2}$ (figure  \ref{F:surf_SF_comp}). Approximate surface values of unity indicate the region in $\Rat$--$E$ space obeys the dissipation-free scaling law (\ref{E:oldNu}) for geostrophic turbulence. It also  implies  $P(E) \equiv 0$ in (\ref{E:Nu}) for the stress-free surface as expected given the absence of Ekman pumping.

\section{No-slip results}

\begin{figure}
\includegraphics[width=.44\textwidth]{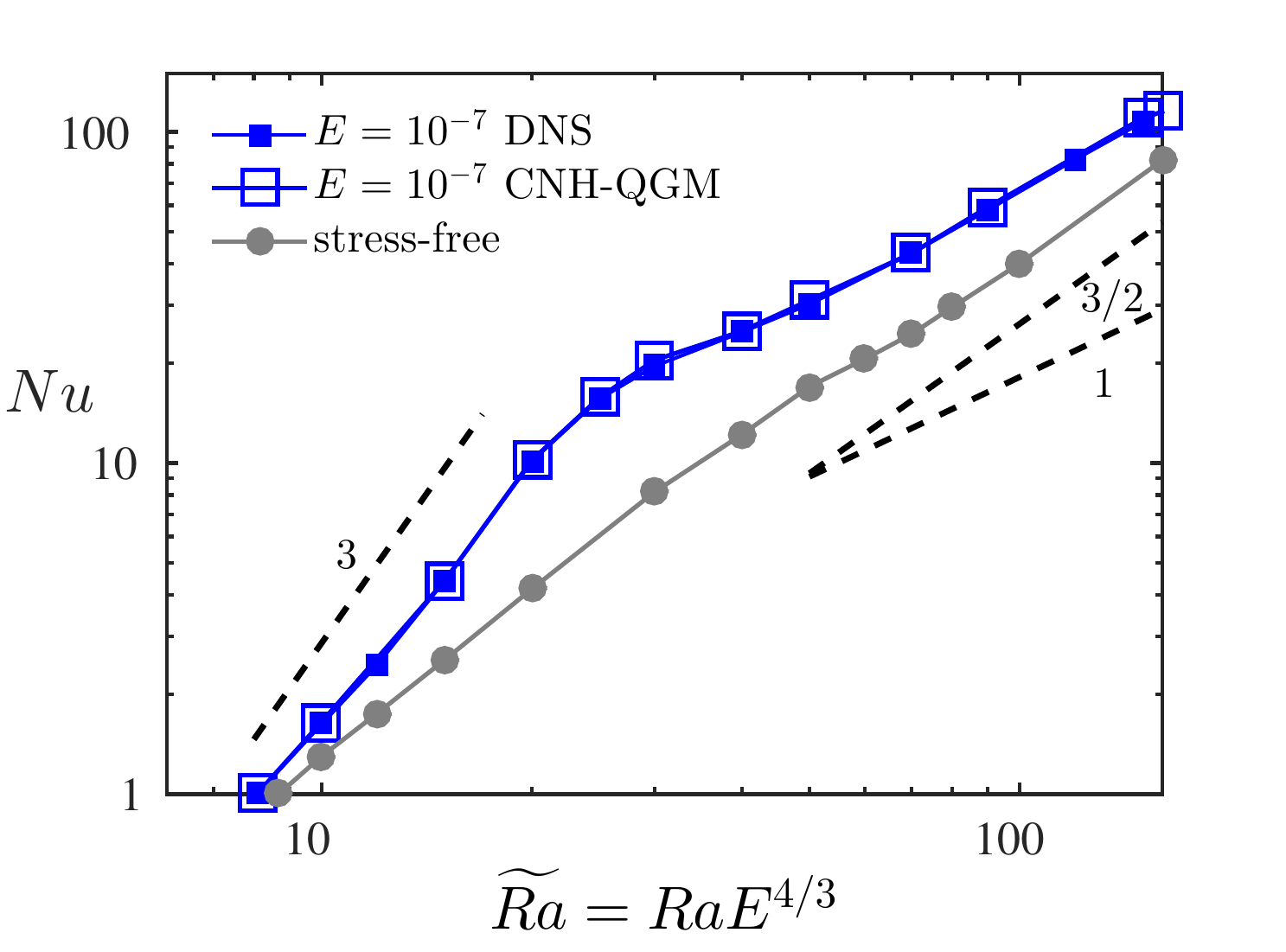}
\caption{ \small  $Nu$ vs $\Rat$ at fixed $E = 10^{-7}$ for DNS (filled squares) and CNH-QGM (open squares). This curve represents the lower bound of achievable results for DNS and upper bound for the CNH-QGM. 
For comparison stress-free results are included (circles). This figure has been reproduced from \cite{Plumley} with an extended range of $\Rat$.}
\label{F:RAoverlap}
\end{figure}

\begin{figure}[h]
\includegraphics[width=.42\textwidth]{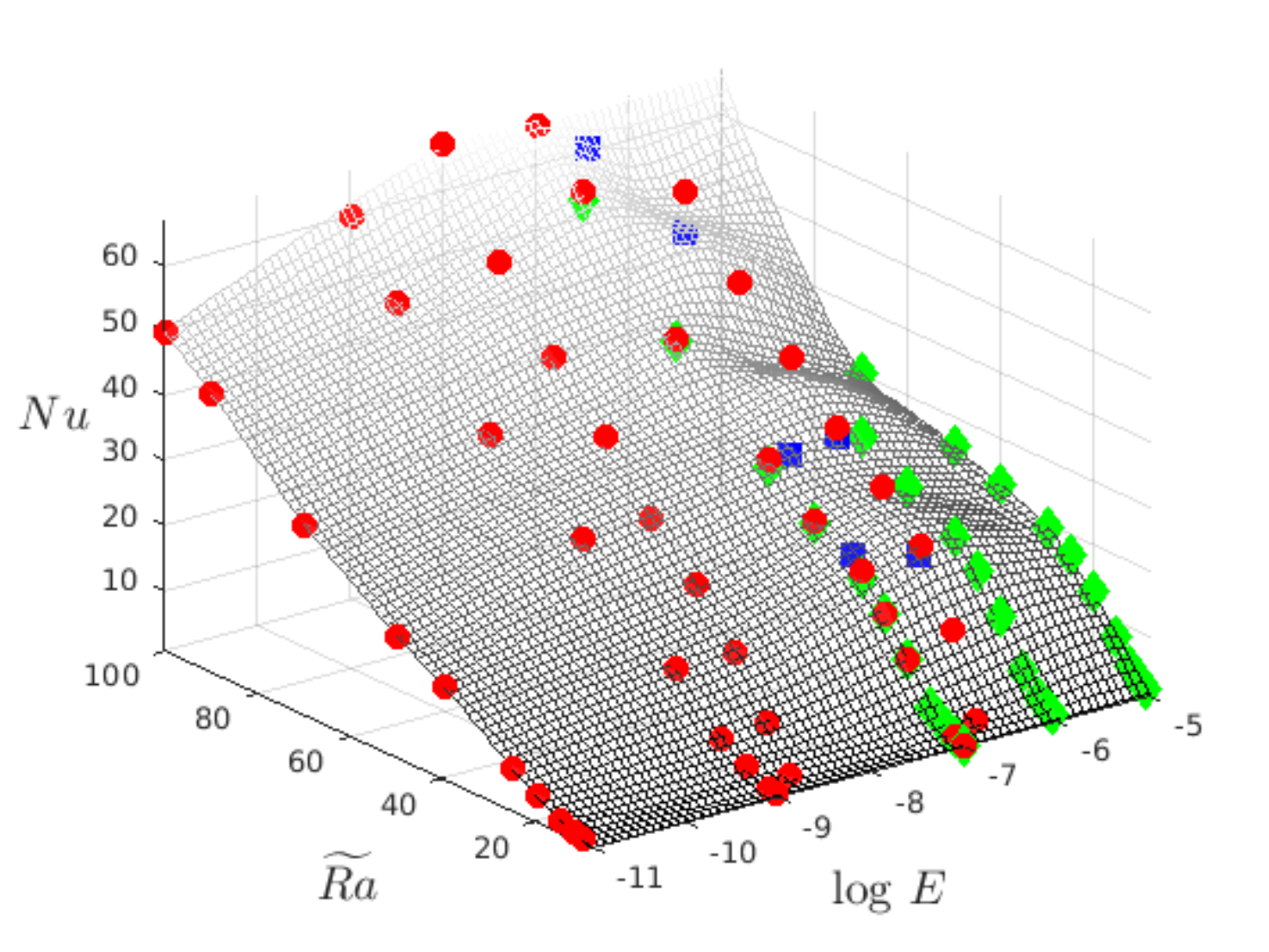} 
\caption{\small 
2D Surface plot of $Nu$ as a function of $\Rat$ and $E$  for no-slip boundaries and $Pr = 1$.
Results marked by (red) circles denote data obtained for $E \le10^{-7}$ in the CNH-QGM.  DNS data for $E \ge10^{-7}$ are obtained from holding $Ra$ fixed (blue squares, \cite{Kunnen16}) and $E$ fixed (green diamonds, \cite{Stellmach}).
}
\label{F:surf}
\end{figure}

The presence of no-slip boundaries in low Rossby-number convection requires the attenuation of the interior velocity field to zero within
an $\mathcal{O}(E^{1/2} H)$ Ekman boundary layer \cite{Greenspan, Julien16}. Within this layer geostrophic balance is lost and a relaxation of the associated horizontal non-divergence of the axial vortical field must occur.  This results in Ekman pumping (and suction) for local cyclonic (anticyclonic) motions. We refer to this phenomena
generically as Ekman pumping where the strength of the dimensional vertical pumping velocities, $w^*_E$, can be related directly to the local vortical motions, $\zeta^*$, via  a classical linear boundary layer analysis, i.e., $w^*_E \sim E^{1/2} H \zeta^*$
\cite{Greenspan}. It has been shown \cite{Julien16} that despite
the $E^{1/2}$ dependence, there exists a transitional $\Rat_t\sim E^{-1/9}$ where vortical motions intensify sufficiently so pumping remains finite in the limit of rapid rotation, i.e.,  $\lim_{E\downarrow 0} w_E\ne 0$. Thus comparative differences with the stress-free case are to be expected.  One example is the enhancement of the convective heat flux and large increases in the heat transport. The increase in heat transport has been investigated and confirmed by DNS at $E \geq 10^{-7}$ \cite{Stellmach} and by the CNH-QGM for $E \leq 10^{-7}$ \cite{Plumley, Julien16} for fixed $E$ explorations.   The quantitative overlap of results at $E = 10^{-7}$ (Figure \ref{F:RAoverlap}), which represents a current lower bound for DNS and upper bound for the CNH-QGM, supports the validity of both methods in their respective parameter regions.

\begin{figure}
\includegraphics[width=.44\textwidth]{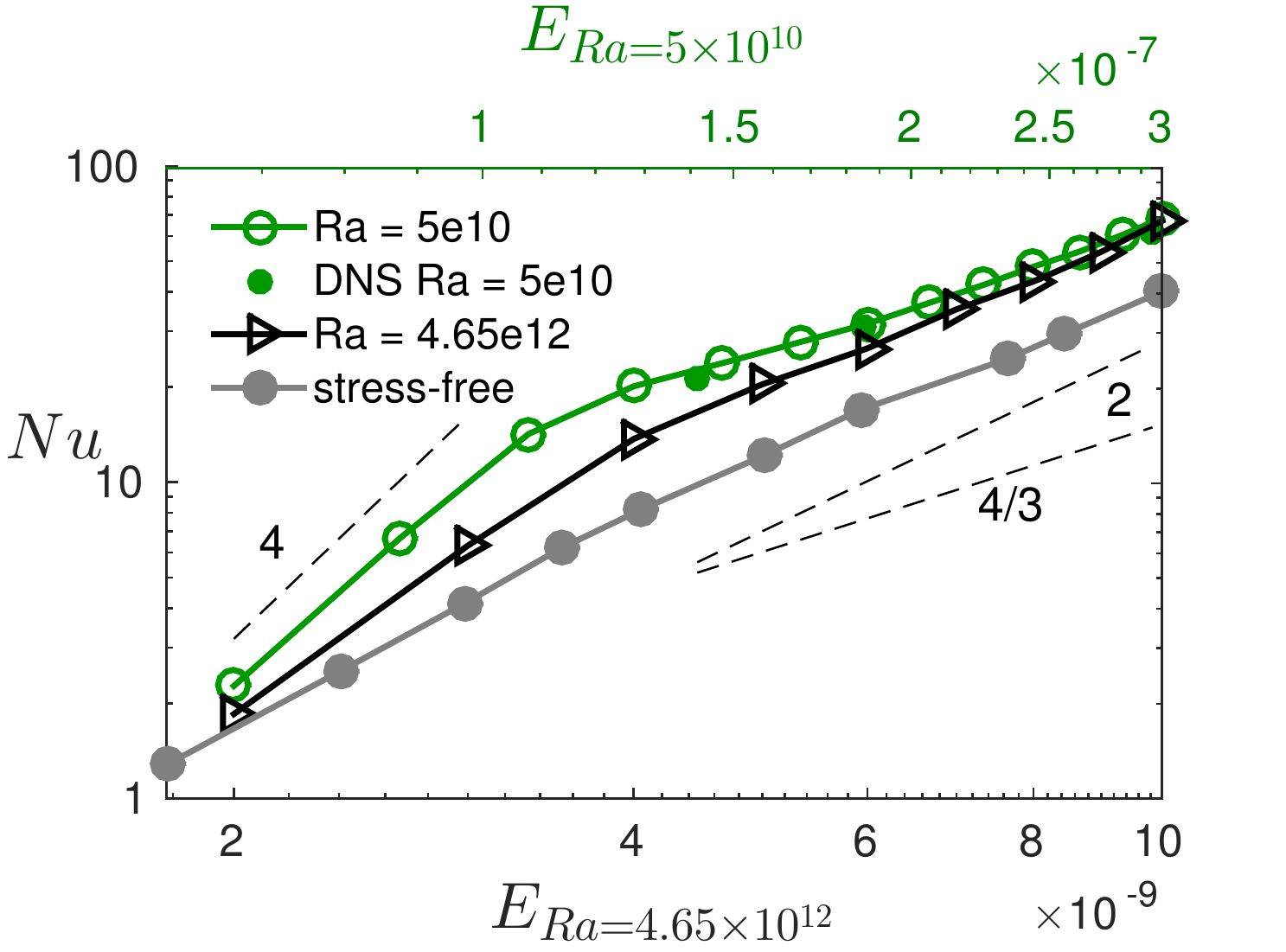} \\ 
\includegraphics[width=.44\textwidth]{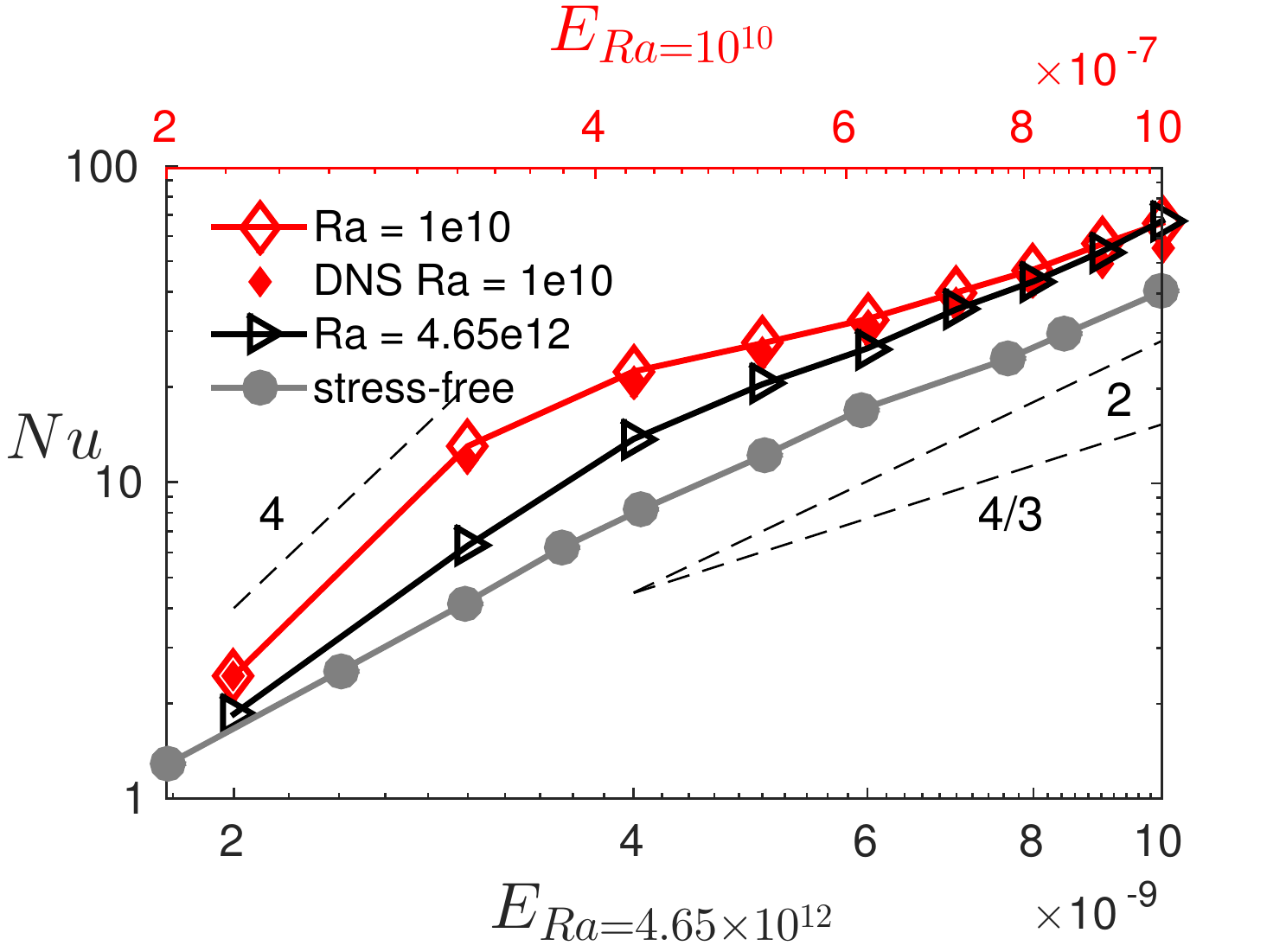}
\caption{ \small 
$Nu$ vs $E$ at fixed $Ra$ for the  CNH-QGM (open symbols) and DNS (filled symbols, \cite{Kunnen16}). Curves are presented for $Ra  = 5 \times 10^{10}$ (green circles in upper plot with top axis), $Ra  = 1 \times 10^{10}$ (red diamonds in lower plot with top axis), and $Ra  = 4.25 \times 10^{12}$ 
(black triangles, bottom axis for upper and lower plots). As $Ra$ increases the slope increases from $\approx 4/3$ to $2$. 
Good quantitative agreement exist between the two approaches. While the stress-free CNH-QGM model does not explicitly contain $E$, the gray stress-free points were calculated by taking each stress-free data point at given $\Rat$ and substituting $Ra = 4.25 \times 10^{12}$ to find the corresponding $E$.
 }
\label{F:overlap}
\end{figure}


Results from DNS \cite{Stellmach, Kunnen16} and the CNH-QGM along different pathways through parameter space are combined to create a surface of the heat transfer $Nu(E, \Rat)$ for the no-slip case. The surface illustrated in figure \ref{F:surf} is a markedly different and more complex surface in comparison with the stress-free surface of figure \ref{F:surf_SF}. 
Specifically, contours at fixed $\Rat$ are no longer parallel to the $E$-axis.  As illustrated in the figure, the effect on the heat transfer is nontrivial even at $E = 10^{-11}$ \cite{Plumley}, due to the addition of Ekman pumping. 
Also evident 
are the much higher $Nu$ slopes in the lower $\Rat$ regime \cite{Stellmach, Plumley} where the cellular morphology occurs \cite{Julien12}. 


\begin{figure}
\includegraphics[width=.44\textwidth]{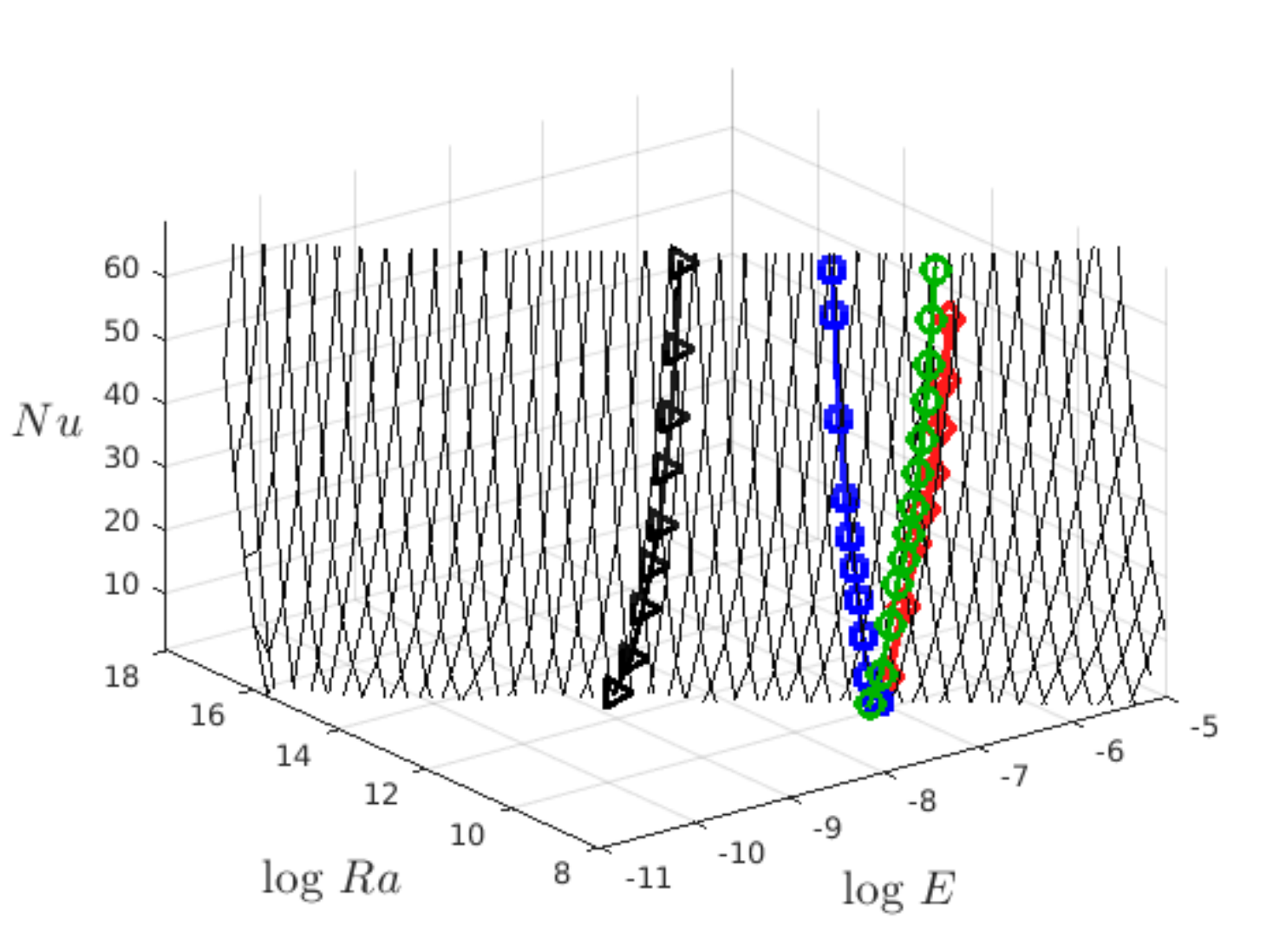} 
\includegraphics[width=.44\textwidth]{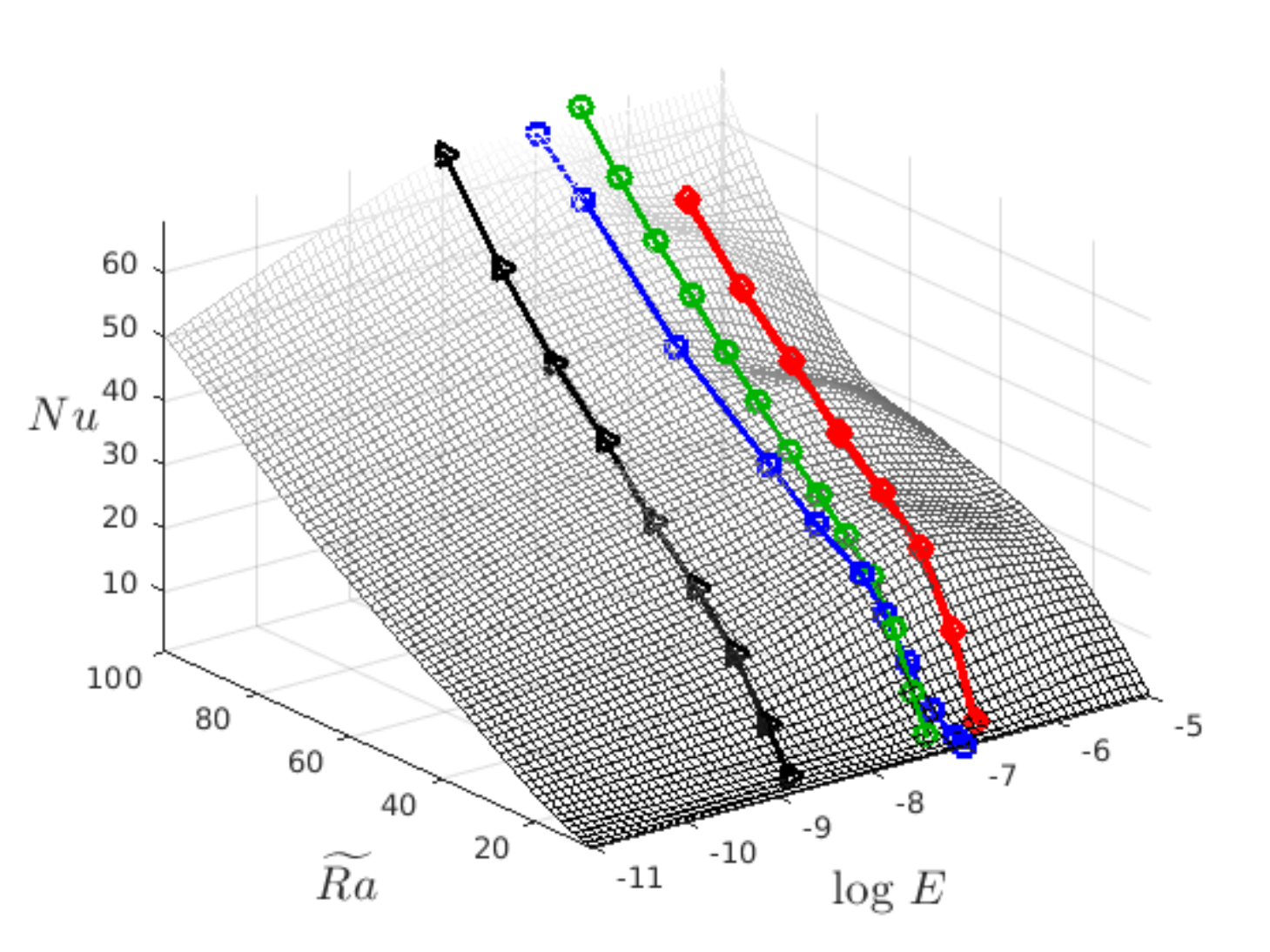}
\caption{ \small 
Surface plots of $Nu(Ra, Ek)$ (top) and $Nu(\Rat, E)$ (bottom) showing the two directions of data used in the surfaces. Colors and symbols agree with the line plots shown in figures \ref{F:RAoverlap} and \ref{F:overlap}. The blue squares show data with constant $E$. The black triangles, red diamonds and green circles show data from constant $Ra$.
 }
\label{F:surf_lines}
\end{figure}

Figure~\ref{F:overlap} shows the heat transfer results for cross-sections through parameter space at fixed $Ra$. The result at $Ra = 5 \times 10^{10}$ (upper plot), with $ E $ near $ 10^{-7}$, shows agreement between DNS \cite{Kunnen16} (3 data points) and the CNH-QGM.  These curves do display a loss of agreement towards the upper range of $E$, which is especially evident for the $Ra = 10^{10}$ data (lower plot). This divergence signals a break down of validity for the CNH-QGM for values near $E = 10^{-6}$. 

The transition of the no-slip curves away from the stress-free curve (gray line, figure~\ref{F:overlap}) can be predicted by $\Rat_t > c E^{-1/9}$ or $E_t > c^{9/13} Ra^{-9/13}$ as found in \citet{Julien16} and fit with $c \approx 1$ \citep{Plumley}. For all of the curves in figure \ref{F:overlap}, the transition occurs below the lowest $E$ plotted; for example, the constant $Ra = 4.65 \times 10^{12}$ data has a predicted transitional $E_t \approx 1.7 \times 10^{-9}$. Within the cellular regime ($\Rat \lesssim 25$), the slopes are much higher than the stress-free slopes, nearing $\alpha \approx 3$ in figure \ref{F:RAoverlap} and equivalently $\beta \approx 4$ in figure \ref{F:overlap} before the transition to the GT regime. This is consistent with recent laboratory and DNS investigations \cite{Cheng}.
The transition from the cellular to GT regimes occurs when the horizontally averaged temperature gradient reaches its minimum and the slope stabilizes at a finite value, indicating unstable stratification in contrast to an isothermal interior \cite{Julien12}.


In figure~\ref{F:surf_lines} the $Nu$ curves obtained at fixed $Ra$ and fixed $E$ (figures~\ref{F:overlap} and \ref{F:RAoverlap} respectively) are illustrated as pathways on the $Nu(Ra,E)$ and $Nu(\Rat, E)$ surfaces.  The directionality of  fixed $Ra$ or $E$  is clear in the $Nu(Ra,E)$ figure, although an asymptotically more useful view is obtained for $Nu(\Rat,E)$, where the constant $Ra$ lines appear as diagonal cross sections. Unlike the stress-free surface, the  complexities extend below $E = 10^{-7}$. 


\begin{figure}
\includegraphics[width=.48\textwidth]{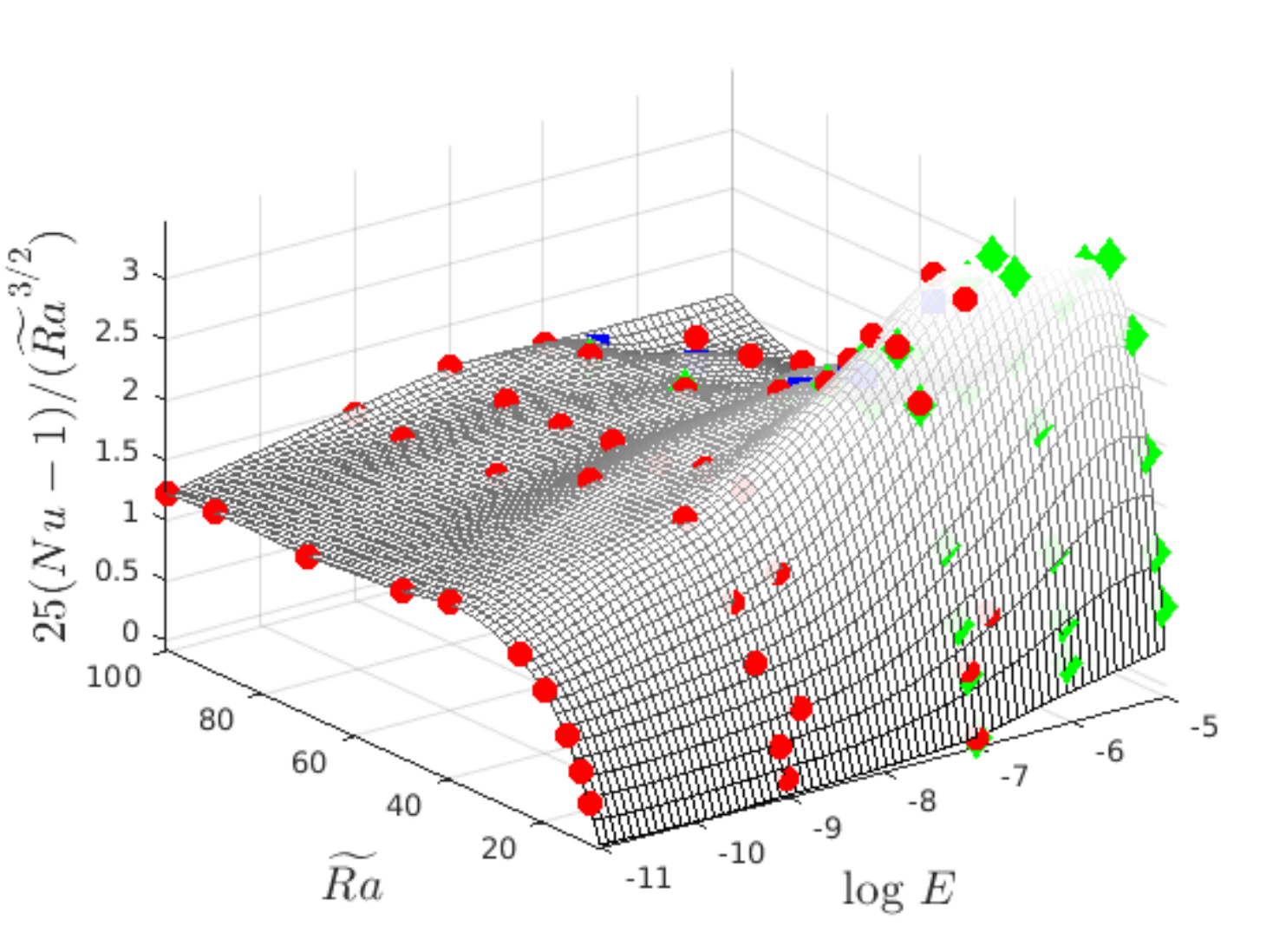} 
\caption{\small 
Compensated 2D surface of $25(Nu-1 )/\Rat^{4/3}$ for no-slip boundaries. 
Data from the CNH-QGM are denoted by (red) circles. DNS data are included for fixed $Ra$ (blue, squares) \cite{Kunnen16} and fixed $E$ (green, diamonds) \cite{Stellmach}.
}
\label{F:surf_compRA}
\end{figure}

\subsection{Quantifying the effects of pumping within the low $E$, GT regime} 
\begin{figure}[h]
\includegraphics[width=.48\textwidth]{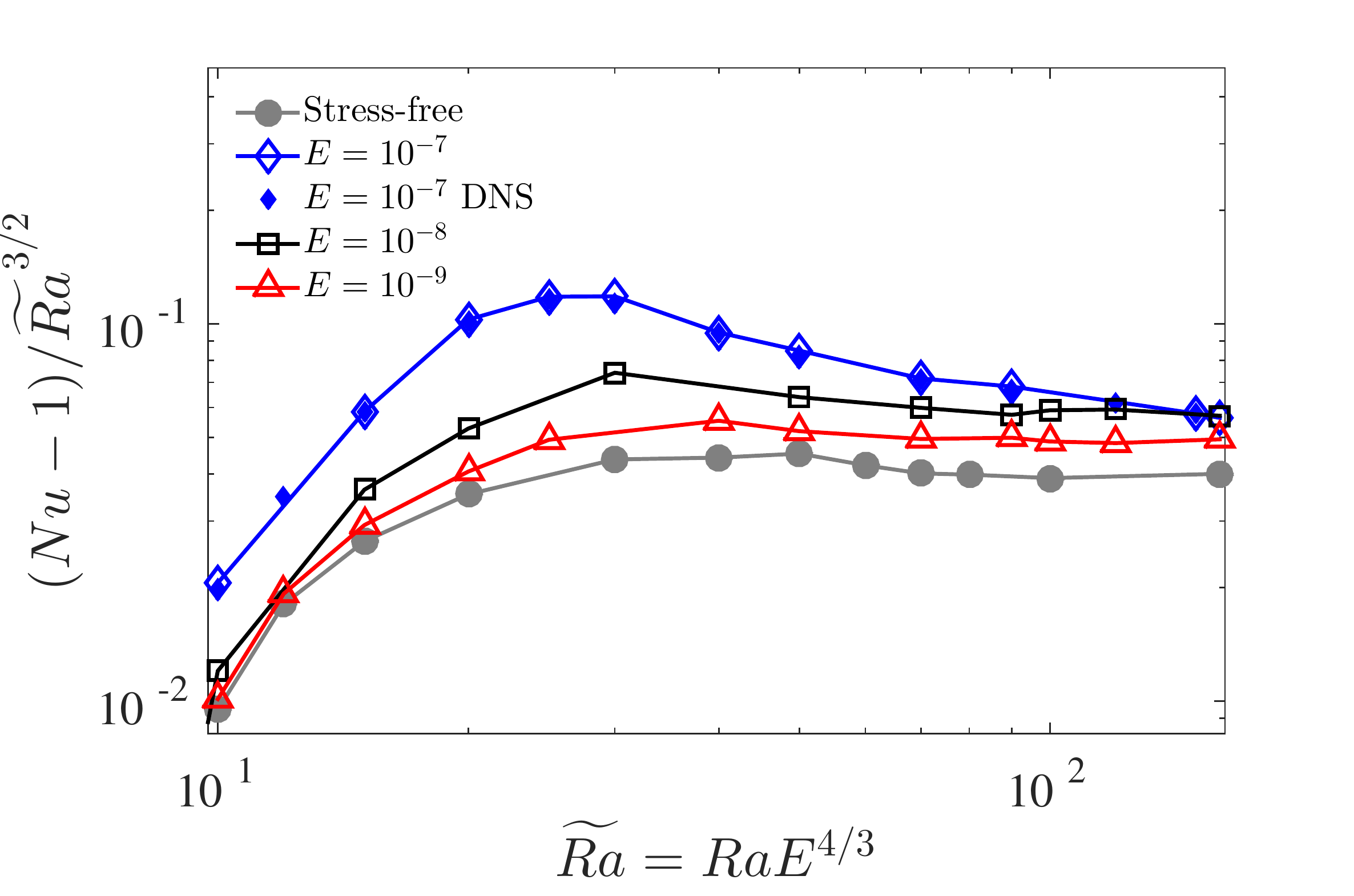} 
\caption{\small Compensated plot of $ (Nu-1)/\Rat^{3/2}$ for the $Pr = 1$ CNH-QGM results for $E = 10^{-7}$ (diamonds), $E = 10^{-9}$ (squares), and $E = 10^{-11}$ (triangles). The stress-free results are plotted as gray circles. }
\label{F:compensated}
\end{figure}

\begin{figure}[h]
\includegraphics[width=.48\textwidth]{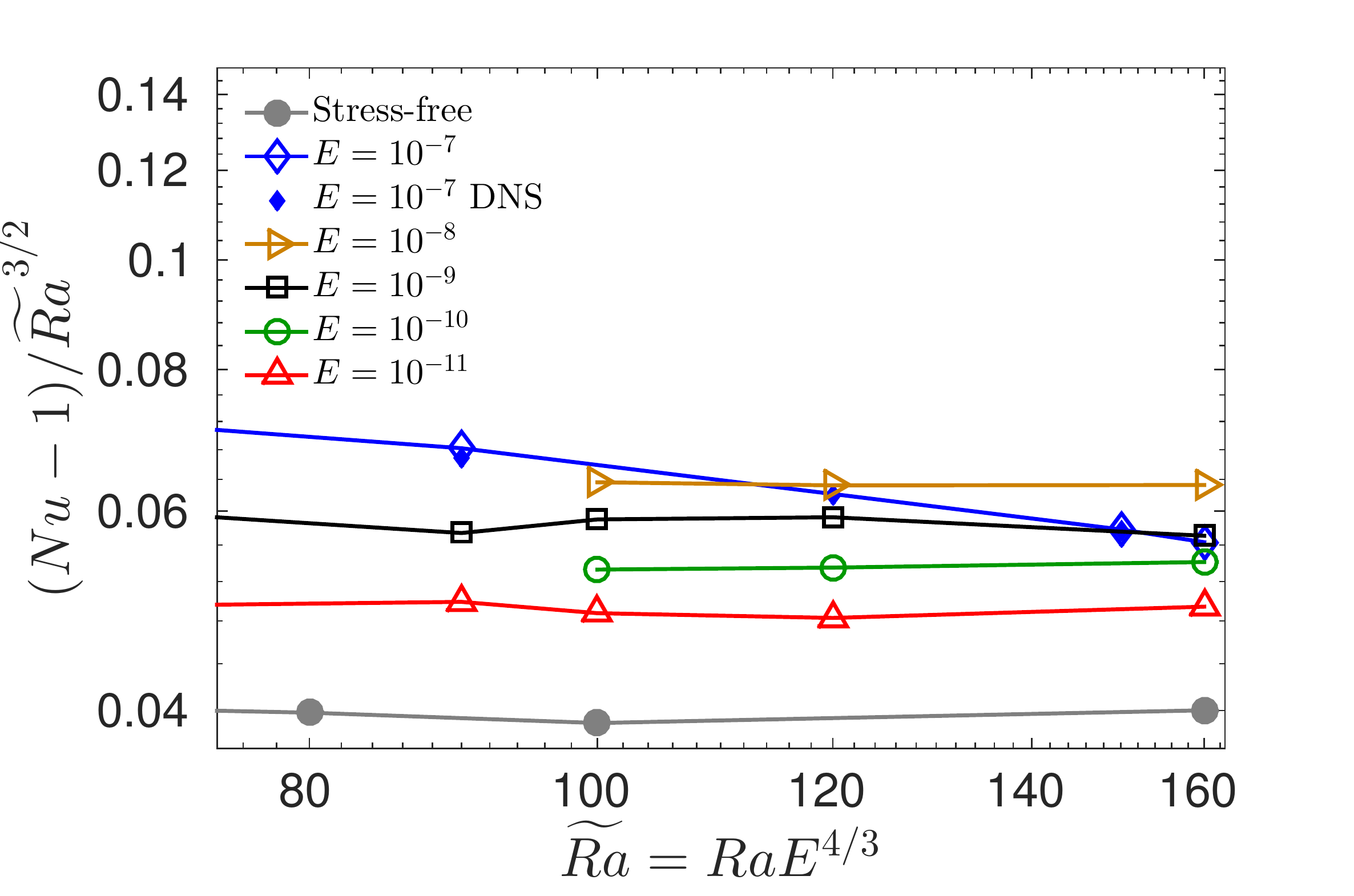} 
\caption{\small  An enlarged section of figure~\ref{F:compensated} for $80 \leq \Rat \leq 160$ and 5 values of $E$. Results include CNH-QGM (open symbols) and DNS (closed) for $E= 10^{-7}$ (diamonds) and CNH-QGM results for $E = 10^{-8}$ (right-facing triangles), $E = 10^{-9}$ (squares), $E = 10^{-10}$ (circles), $E = 10^{-11}$ (upward triangles) and stress-free (filled circles). }
\label{F:compensatedzoom}
\end{figure}


For comparison with the stress-free surface, the no-slip surface is compensated by the dissipation-free results (\ref{E:SFnu2}) in figure~\ref{F:surf_compRA}. This makes explicit the enhancement due to Ekman pumping as the surface appears to be a plane, with height greater than $1$, that is tilted upwards with $E$ in the high $\Rat$ - low $E$ domain. This implies that the stress-free fit is a useful base model for the no-slip curve at high $\Rat$, which only requires alteration to account for the increase in the heat transfer. 

The enhancement can be directly computed from the compensated heat transfer results obtained for cross-sections at multiple values of fixed $E$  as plotted in figure~\ref{F:compensated}. For $E < 10^{-7}$ an important observation is the evolution of the compensated curves to the dissipation-free exponent $\alpha=3/2$ for $\Rat\gtrsim 100$. This is more explicit in a blow up of the region $\Rat>80$  (figure \ref{F:compensatedzoom}).

The $E = 10^{-7}$ data exhibits no such convergence to  $\alpha=3/2$ in this region and continues to decrease with $\Rat$.
Instead we observe that, in addition to multiplicative effects, the ageostrophic effect of Ekman pumping now diminishes the 
 heat transport efficiency to exponent $\alpha \approx 1.2$. The transition to a self-similar scaling regime  (\ref{E:SFnu2}) \ thus occurs for $E < 10^{-7}$.  This elicits a few remarks.  Given that  $E = 10^{-7}$ represents an established quantitative benchmark between DNS and CNH-QGM 
 data this supports an extended range of $\Rat$ for which the CNH-QGM is valid.  Despite this endorsement of the CNH-QGM model, the $E = 10^{-7}$ result complicates the understanding of the enhancement as it appears entering the self-similar regime is associated with non-monotonic behavior in $E$. Indeed, given that CNH-QGM is geostrophically balanced in both interior and thermal boundary layer,  the result $\alpha \approx 1.2$ cannot be attributed to loss of geostrophic balance in the latter. The transition for loss of balance has been predicted in 
 \cite{Julien12PRL} to be given by $\Rat_{t_2} \approx \tilde{d} E^{-4/15}$  (or equivalently $Ra_{t_2} \approx  d E^{-8/5}$). Thus, the 
 exploration of parameter space $\Rat\le 160 < \Rat_{t_2} $ implies $\tilde{d} \gtrsim 2.17$. It then appears that the observed non-monotonicity,
captured by the DNS and CNH-QGM,  can be attributed solely to the ageostrophic effects of the Ekman pumping and the associated Ekman boundary layer. 

The influence of pumping on the heat transfer for $E \approx 10^{-7}$ is also visible in figure \ref{F:overlap} for the curves with constant $Ra = 5 \times 10^{10}$  (green circles) and $Ra= 10^{10}$ (red diamonds). These curves show $\beta \approx 4/3$, demonstrating that the heat transfer slope is similarly diminished from the stress-free result (of $\beta = 2$) as for the constant $E = 10^{-7}$ data. 
However, the curve at higher $Ra = 4.65 \times 10^{12}$ ($E \approx 10^{-9}$) reaches the self-similar regime and runs parallel to the stress-free result with $\beta \approx 2$.

This can be understood as a Rossby number effect. Departures from the stress-free scaling exponents are expected for larger $Ro \gtrsim .03 $. For the three curves in figure \ref{F:overlap}, the Rossby number ranges within the GT regime are $ .05 \leq Ro_{GT} \leq .1$ for the $Ra = 10^{10}$ data (red diamonds), $.034 \leq Ro_{GT} \leq .067$ for the $Ra = 5 \times 10^{10}$ data (green circles) and $.011 \leq Ro_{GT} \leq .022$ for the $Ra = 4.65 \times 10^{12}$ data (black triangles). The Rossby number acts as a control parameter and for high $Ro$ the results depart from the expected scaling behavior. This is confirmed in figure \ref{F:compensatedzoom} as the curves follow the expected scaling behavior for the runs with lower $Ro$ (lower $E$). For the $E = 10^{-7}$ data at $\Rat = 160$, $Ro = .058$, whereas for $E = 10^{-8}$ and $\Rat = 160$, $Ro = .027$, and only the $E\leq10^{-8}$ -- lower $Ro$ curves result in the self-similar solution. 
  
%

\begin{figure}[h]
\includegraphics[width=.48\textwidth]{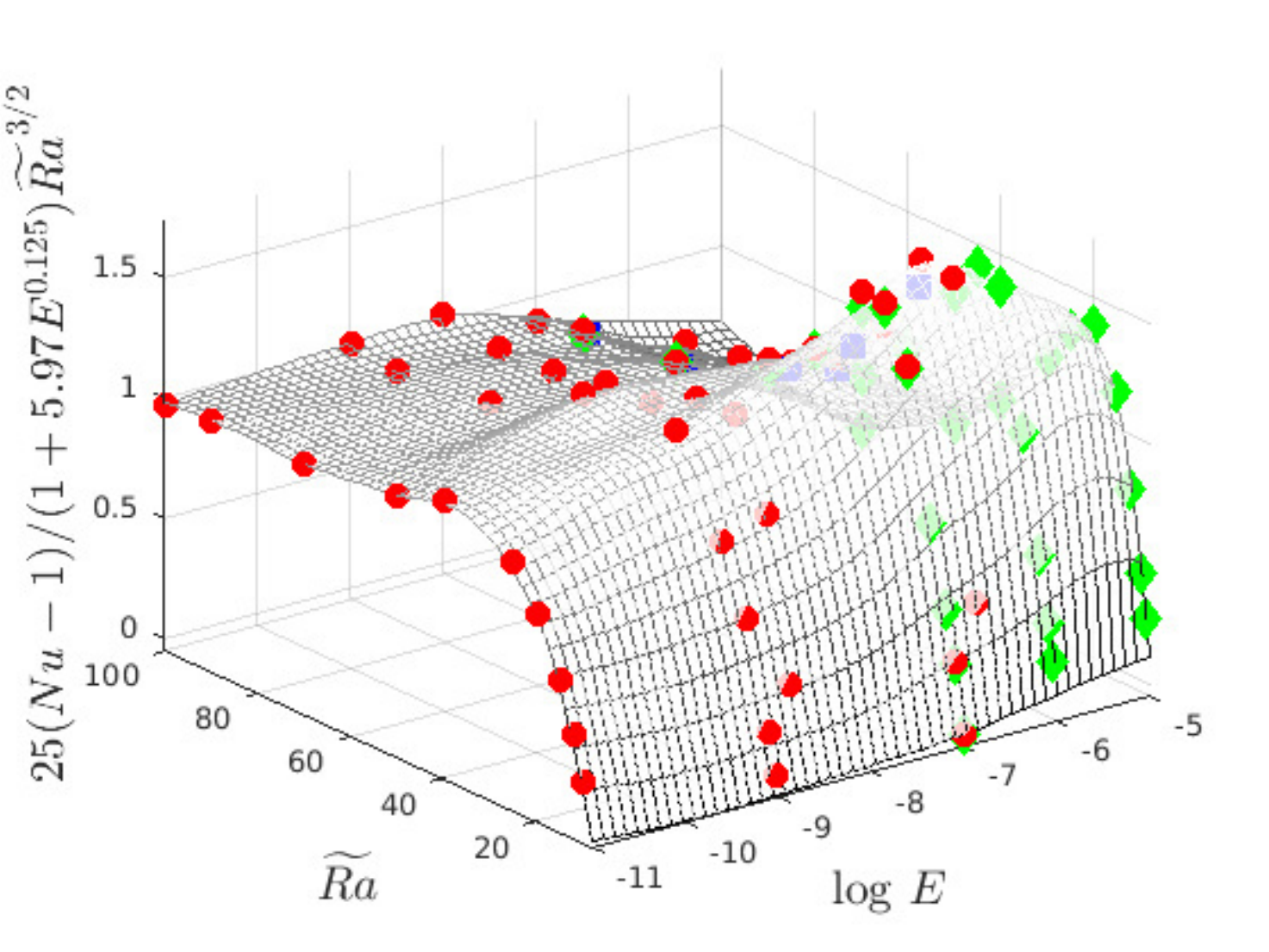} 
\caption{\small Compensated 2D surface of $Nu$--$\Rat$--$E$ using (\ref{E:nu_NS_fit}).
Data from the CNH-QGM are denoted by (red) circles. DNS data are included for fixed $Ra$ (blue, squares) \cite{Kunnen16} and fixed $E$ (green, diamonds) \cite{Stellmach}. While the high $E$ -- low $\Rat$ shows variation, the fitting applies to the low $E$ -- high $\Rat$ data and the effectiveness of the fit can be seen in the flattening of the surface to a value of 1 in this region. 
}
\label{F:surf_comp}
\end{figure}

The combination of observations from figures~\ref{F:surf}, \ref{F:surf_compRA}, and  \ref{F:compensatedzoom} leads to (\ref{E:Nu}), where $P(E)$ accounts for the vertically shifted increase and the decreasing impact of Ekman pumping as $E \downarrow 0$. Therefore, we propose $P(E) = c_2 E^{\delta}$ for the self-similar regime. The values of $c_2$ and $\delta$ can be fitted from the values of the $E$ curves in figure \ref{F:compensatedzoom}. Additional points for $E = 10^{-8}$ and $10^{-10}$ were calculated to supply four points to fit, and the $E = 10^{-7}$ was not included.  Using the increase for each in relation to the stress-free value at $\Rat = 160$, $\delta = .126 \pm .012 \approx 1/8$ and $c_2 = 5.97 \pm 1.31$ provides an empirical estimate of the expected increase due to pumping within the GT regime.  Thus, for $Pr = 1 $,

\be
Nu -1 = \frac{1}{25} \, (1 + 5.97 \, E^{1/8}) \, \Rat^{3/2} \, .
\label{E:nu_NS_fit}
\ee

From the viewpoint of probing the geostrophic regime by decreasing $E$, the $1/8$ exponent indicates 
a slow decrease in the effect of Ekman pumping. Indeed, given the amplification factor $5.97 \, E^{1/8}$, a relative difference of 
$10^{-1}$ requires $E\ge 10^{-15}$. This is a bound satisfied by rotating convection in  most planetary and stellar objects \cite{jmA15}.



Figure \ref{F:surf_comp} shows the surface compensated according to  (\ref{E:nu_NS_fit}). The high $\Rat$ values compress to a compensated value $\approx 1$, showing that the fit captures the expected curves well in this region. The low $\Rat$, high $E$ still shows variations in the surface. This region is characterized by cellular motions rather than GT and is not expressed as well in the fit.


\begin{figure}
\includegraphics[width=.44\textwidth]{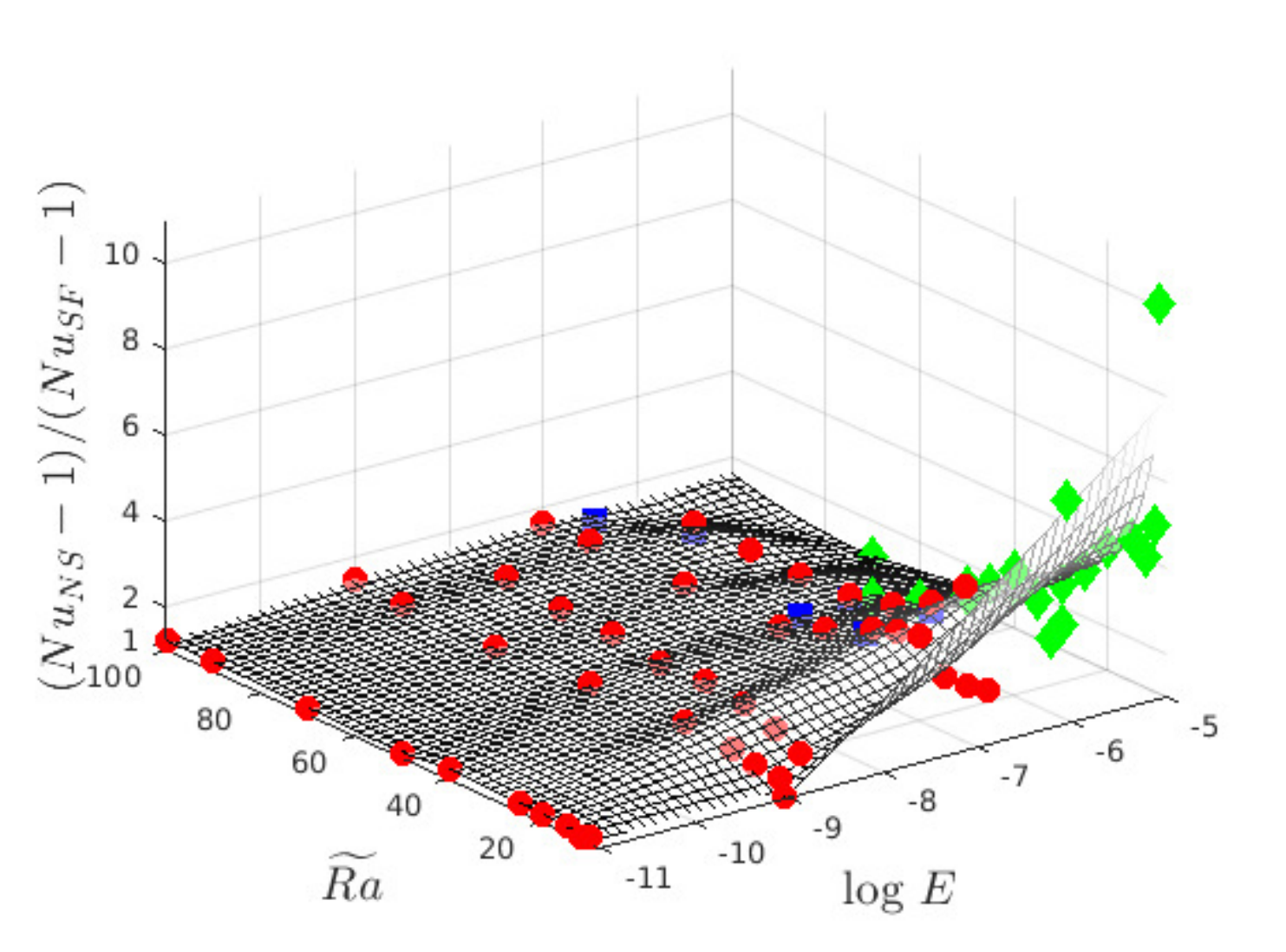} 
\caption{\small  The 2D surface of $(Nu_{NS}-1)/(Nu_{SF}-1)$. Here the subscripts denote either no-slip (NS) or stress-free (SF) $Nu$ values. }
\label{F:surf_NS-SF2}
\end{figure}

Another way to examine the no-slip results is to ignore the scaling laws and instead compare the ratio of $(Nu_{NS}(E,\Rat) - 1)/(Nu_{SF}(E,\Rat)-1)$, where the subscripts denote either no-slip (NS) or stress-free (SF) $Nu$ values. Figure \ref{F:surf_NS-SF2} shows the results of this calculation. The majority of the plot shows that this ratio is bounded by 2 and the surface is relatively flat. 
This simultaneously indicates a bound on the enhancement due to Ekman pumping and highlights the  similarity in the efficiency of heat transport as measured by the exponents for both the stress-free and no-slip cases. 
The strong effect of Ekman pumping within the cellular regime is also observed in this figure in the low $\Rat$, high $E$ domain.
%

\section{Conclusions}

\begin{figure}[b]
\includegraphics[width=.44\textwidth]{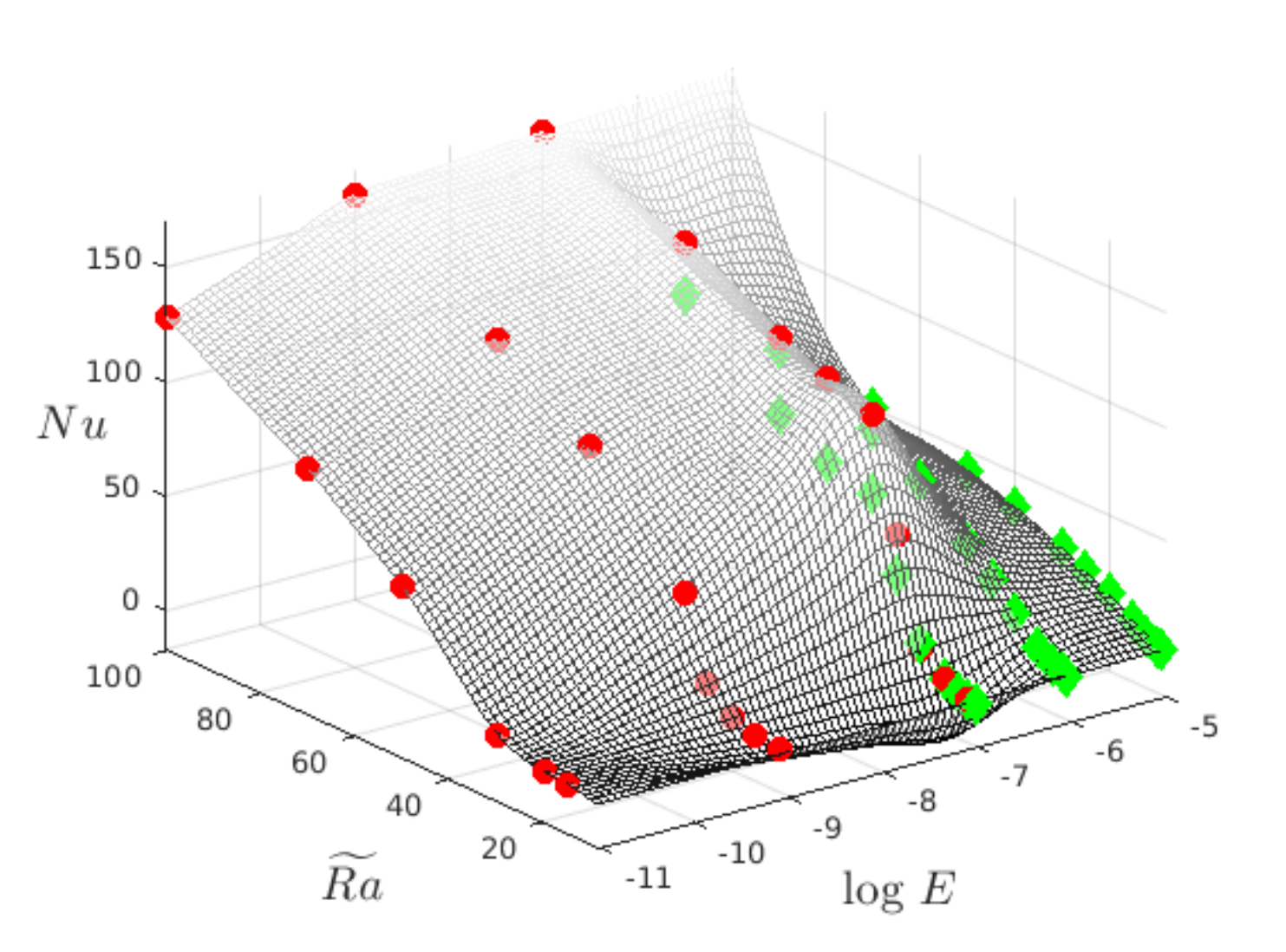}
\includegraphics[width=.44\textwidth]{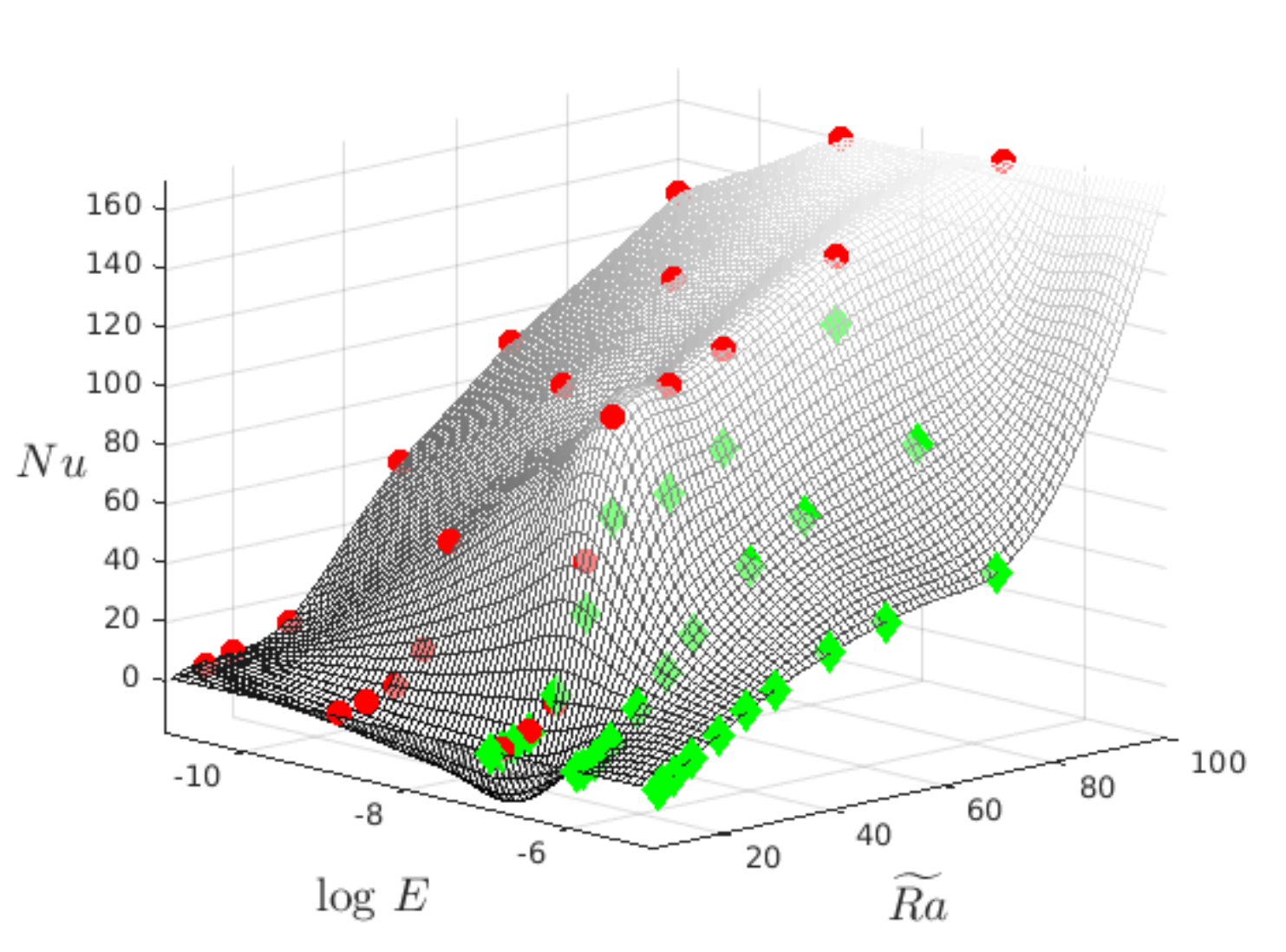}  
\caption{\small 
Two perspective surface plots of $Nu$--$\Rat$--$E$ for no-slip boundaries and $Pr = 7$.
Data from the CNH-QGM are denoted by (red) circles and DNS results obtained at constant $E$ \cite{Stellmach} are denoted by (green) diamonds.}
\label{F:surf_NS-SF3}
\end{figure}
%

\begin{figure}
\includegraphics[width=.48\textwidth]{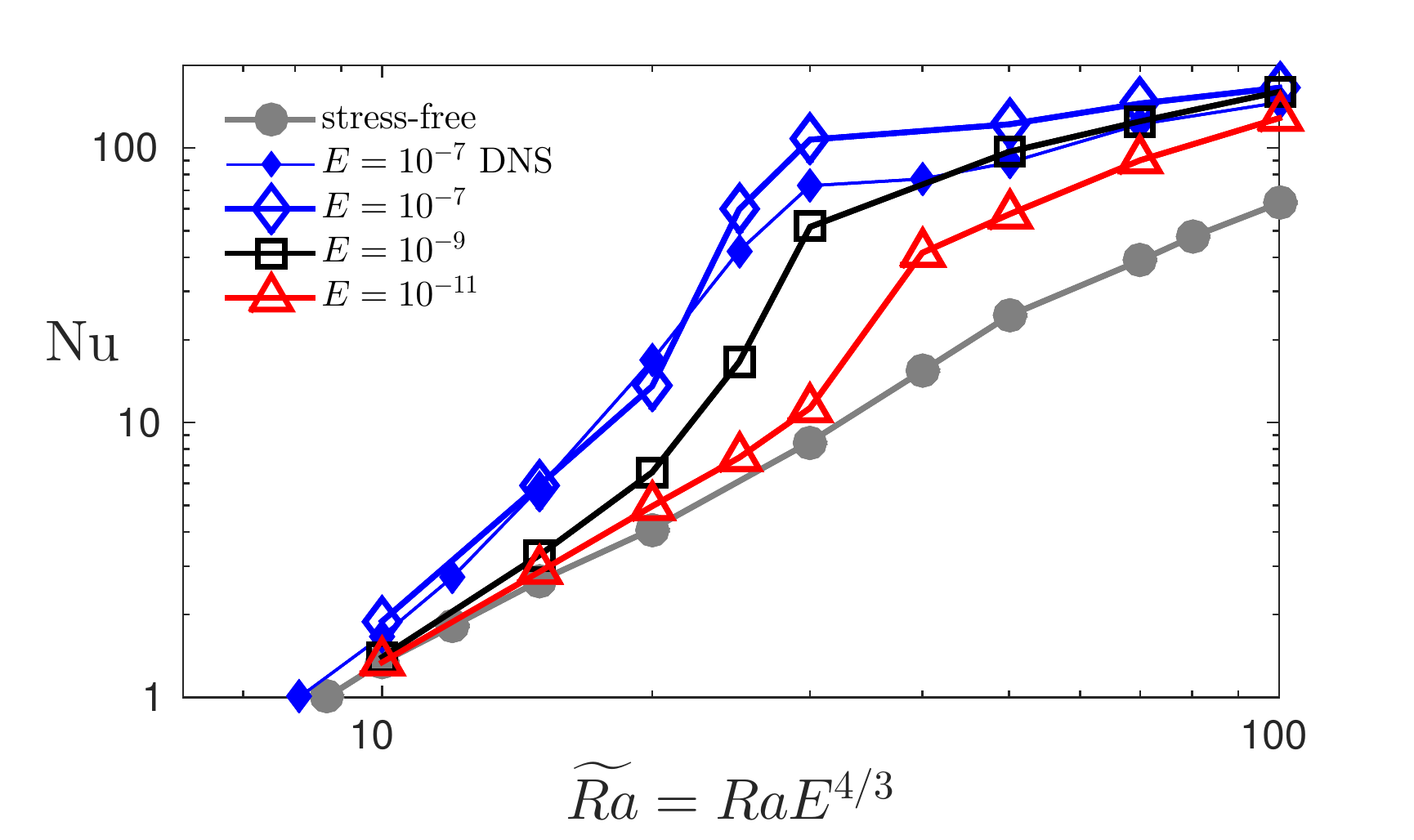} 
\caption{\small $Nu$ vs $\Rat$ for fixed $E = 10^{-7}$ (diamonds), $E = 10^{-9}$ (squares), $E = 10^{-11}$ (triangles) and the stress-free (circles) obtained by the CNH-QGM for $Pr=7$. The DNS data at $E = 10^{-7}$ are included as filled diamonds.}
\label{F:lowEPr7}
\end{figure}

The heat transfer scalings as function of thermal forcing and rotation rate are a primary focus of rotating convection studies.
However thermal and mechanical difficulties, and issues of numerical stability, limit laboratory experiments and DNS to $E\ge 10^{-7}$. 
This value is known to be on the boundary for probing geostrophic dynamics \cite{Ecke14}. Different approaches for entering the geostrophic regime (fixed $Ra$ or $E$) have also produced results that appear incompatible. We have demonstrated that the $Nu(E,\Rat)$
surface is complex and that by mapping the full 2D surface of results the aforementioned incompatibility can be understood as cross-sections through the surface in different directions. 

Indeed,  the surfaces show that any observed scaling exponent is dependent on the particular cross-section through $\Rat$-$E$ parameter space. The surface highlights the vertical shift of the heat transport results from the stress-free to no-slip cases as well as the additional $E$ dependence caused by pumping. This dependence creates a more convoluted surface as the heat transfer is a non-monotonic function of $E$ even within the geostrophic regime.
The deviation from the simplified 1D stress-free slope emerges from the $E$ dependence of the Ekman pumping velocity for no-slip boundaries. Thus, varying $E$  at fixed $Ra$ for the no-slip case continuously varies the strength of pumping and adds greater complexity in understanding the results despite the ease of experimental design.

For low $E \leq 10^{-8}$ and high $\Rat \geq 80$, the results reach an utlimate scaling that is used to characterize this section of parameter space.  Utilizing the stress-free scaling law as a guide and fitting the heat transfer enhancement due to pumping, the scaling law $Nu -1 = (1/25) \, (1 + 5.97 \, E^{1/8}) \, \Rat^{3/2}$ applies for the $Pr =1 $ no-slip case.  

We finally note that a majority of laboratory experiments for RRBC are performed for water where $Pr\approx 7$.
Mapping the surface for $Pr =7$ (or other $Pr \ge 3$) is complicated by a lack of laboratory and numerical data in the geostrophic 
regime, and a reduced overlap of results at $E = 10^{-7}$ between DNS and the CNH-QGM \cite{Plumley}. The difficulties are compounded by the 
 extended range of parameter space dominated by columnar or plume structures for $Pr \geq 3$ \cite{Sprague}. Such coherent structures act as conduits for efficient heat transfer and most likely do not reflect the ultimate characteristics of geostrophic turbulence that has been shown to throttle the heat flux. 
The present status of the 2D surface is illustrated in figure~\ref{F:surf_NS-SF3}.  The increased magnitude of $Nu$ resulting from the columnar morphology is evident in the diagram when compared to the $Pr=1$ results in figure~\ref{F:surf_SF}. However, we note that after a steep increase in the $Nu$-$\Rat$ curve at fixed $E$ the slope appears to be settling into a regime that bears similarity with the dissipation-free regime observed
for $Pr=1$ (Figure~\ref{F:lowEPr7}). Future results by the UCLA Spinlab, Eindhoven TROCONVEX, and Gottingen Uboot laboratory experiments should help by adding data at higher $\Rat$ to this surface for $Pr = 7$.

\section*{Acknowledgments} 
This work was supported by NASA Headquarters under the NASA Earth and Space Science Fellowship Program (M.P.) and the National Science Foundation under EAR grants \#1320991 \& CSEDI \#1067944 (K.J., P.M.).  S.S. gratefully acknowledges the Gauss Centre for
Supercomputing (GCS) for providing computing time through the John von Neumann
Institute for Computing (NIC) on the GCS share of the supercomputer JUQUEEN at
Jülich Supercomputing Centre (JSC) in Germany. This work utilized the NASA Pleiades supercomputer and the Janus supercomputer, which is supported by the National Science Foundation (award number CNS-0821794) and the University of Colorado Boulder.


%

\end{document}